\documentclass{elsart}

\usepackage[dvips]{graphicx}
\usepackage{amssymb}
\usepackage{xspace}

\def\lep{{\sc lep}\xspace}

\def\lepb{{\sc lep2}\xspace}

\def\mc{{\sc mc}\xspace}
\def\csf{{\sc sf}\xspace}
\def\is{{\sc is}\xspace}
\def\fs{{\sc fs}\xspace}
\def\ifl{{\sc ifl}\xspace}
\def\isr{{\sc isr}\xspace}

\def\qed{{\sc qed}\xspace}
\def\qcd{{\sc qcd}\xspace}
\def\es{{\sc es}\xspace}

\def\lls{{\sc LL}\xspace}

\def\alp{{\tt ALPHA}\xspace}

\def\cern{{\sc cern}\xspace}

\begin{document}

\newlength{\caheight}
\setlength{\caheight}{12pt}
\multiply\caheight by 7
\newlength{\secondpar}
\setlength{\secondpar}{\hsize}
\divide\secondpar by 3
\newlength{\firstpar}
\setlength{\firstpar}{\secondpar}
\multiply\firstpar by 2

\hfill
\parbox[0pt][\caheight][t]{\secondpar}{
  \rightline
  {\tt \shortstack[l]{
      FNT/T-2000/08
      }}
  }

\begin{frontmatter}

\title{Higher-order QED corrections to single-$W$ production 
in electron-positron collisions}

\author[pavia]{G.Montagna}
\author[ferrara]{M.Moretti}
\author[pavia1]{O.Nicrosini}
\author[pavia]{A.Pallavicini}
\author[pavia1]{F.Piccinini}

\address[pavia]{Dipartimento di Fisica Nucleare e Teorica, 
Universit\`a di Pavia\\
  and INFN, Sezione di Pavia, via A. Bassi 6, 27100, Pavia, Italy}
\address[pavia1]{INFN, Sezione di Pavia and 
Dipartimento di Fisica Nucleare e Teorica, \\ Universit\`a di Pavia, 
via A. Bassi 6, 27100, Pavia, Italy}
\address[ferrara]{Dipartimento di Fisica, Universit\`a di Ferrara\\
  and INFN, Sezione di Ferrara, Ferrara, Italy}

\begin{abstract}
Four-fermion processes with a particle lost in the beam pipe
are studied at \lep to perform precision tests of the electroweak 
theory. Leading higher-order \qed corrections to such 
processes are analyzed within the framework of the Structure Functions 
(\csf)
approach. The energy scale entering the \qed \csf is determined 
by inspection of the soft and collinear limit of 
the $O(\alpha)$ radiative corrections to the four-fermion final states,
paying particular attention to the process of single-$W$ production. 
Numerical predictions are shown in realistic 
situations for \lep experiments and compared with existing results. 
 A Monte Carlo event generator, including exact tree-level matrix 
elements, vacuum polarization, higher-order leading \qed 
corrections and anomalous trilinear gauge couplings, 
 is presented. 
\end{abstract}


\end{frontmatter}

\newpage

\section{\label{intro} Introduction}

Four-fermion final states are of special interest for the physics 
programme 
of \lepb and future high-energy 
electron-positron colliders, being entangled with electroweak 
gauge boson production and decay~\cite{lepyr}.
In particular, the process considered in the present paper, {\it i.e.}

\begin{equation}
\label{wwe}
  e^+ e^- \rightarrow e^- (e^+) \bar \nu_e (\nu_e) \, q^\prime \bar q 
\end{equation}

is peculiar among all the possible four-fermion final states 
because the bulk of its cross section 
is due to two sub-processes, {\it i.e.} $W$-boson pair production and decay

\begin{equation}
\label{wws}
  e^+ e^- \rightarrow W^*  W^* \rightarrow 4 \ {\rm fermions}
\end{equation}

and the radiation of an almost on shell $t$-channel photon from
the electron (positron), with subsequent production of a $W$-boson
and a neutrino

\begin{equation}
\label{sw}
  e^+ e^- \rightarrow \gamma^*  e^+ (e^-) \rightarrow W^* \bar \nu_e (e^-)
  \rightarrow 4 \ {\rm fermions}
\end{equation}

Despite, strictly speaking, both sub-processes (\ref{wws},\ref{sw})
always occur simultaneously and are indistinguishable, channel (\ref{wws}) 
dominates if the electron is emitted at large angle, whereas channel (\ref{sw})
dominates if the electron is emitted in the very forward region,
because of the presence of a quasi-real $t$-channel photon. 

In this paper the process (\ref{sw}) will be addressed, 
by restricting the analysis to the kinematical range 
of forwardly emitted electrons.
This signal is usually referred to as single-$W$ production, since
only the two final jets are detected~\cite{tanaka}.

The importance of this process has been emphasized since long time.
In the \lepb energy range it is fundamental in order to study the 
self-interaction of the gauge bosons, 
together with the process (\ref{wws}),
whereas in the energy regime of future colliders at the 
TeV scale it becomes the dominant
electroweak process. In refs.~\cite{realw,wwrealw} cross sections and 
distributions were calculated in the approximation of real $W$-boson production, 
either
by \cite{realw} studying the reaction $e^+e^-\rightarrow e^- \bar\nu_e W^+$, 
or by \cite{wwrealw,trig} employing the Weizs\"acker-Williams \cite{ww}
equivalent-photon approximation for the $t$-channel photon.
In refs.~\cite{realw,wwrealw,trig} it was pointed out the 
relevance of this
process for the study of trilinear gauge boson couplings and
some assessment of the  sensitivity has been given.
The first full four-fermion calculation, including the crucial effect of 
fermion masses, has been presented in ref.~\cite{grace}, where
the \lepb sensitivity to anomalous gauge couplings has been studied.
Since then, other complete four-fermion calculations of 
the single-$W$ 
process have appeared
in the literature and implemented in computational tools for data 
analysis~\cite{grace,ifl,passarino,dub,bpp}. 
In most of these calculations the effect of 
fermion masses is exactly accounted for 
in the dynamics and kinematics for the whole 
four-fermion phase-space \cite{grace,ifl,dub,bpp}, 
while in the approach of ref.~\cite{passarino}  
the Weizs\"acker-Williams approximation is employed in the 
very forward, collinear region and matched with a massless 
four-fermion computation outside it. 
An up-to-date inventory of the present theoretical status is 
under preparation by the four-fermion working group of the 
\lepb \mc workshop at \cern~\cite{proposal}.   

Measurements of the single-$W$ cross section and the corresponding bounds on
anomalous gauge couplings have been recently 
reported by the \lep collaborations
\cite{tanaka}. Because the foreseen accuracy of final \lepb data 
is of the order of $1$-$2\%$ \cite{proposal}, accurate 
theoretical predictions
 for cross section and distributions are required.

The calculation of the cross section for single-$W$ processes poses several
non-trivial theoretical problems \cite{proposal}. For a realistic account of gauge bosons
properties it is mandatory to include the gauge boson width in the propagator.
In general this mixes a fixed order calculation with an all order resummation
of a class of Feynman diagrams and introduces a violation of the Ward identities
of the theory. This issue is of special importance here since, due to the
$t$-channel photon exchange, even a tiny violation of \qed Ward identities
is enhanced by a factor of $s/m_e^2$. This is indeed the case if
a running width is used in the calculation. This problem has been extensively
studied \cite{fudge,fwidth,floop}, and several options to address it have been
explored. The most theoretically appealing procedure is the fermion loop 
scheme~\cite{fwidth,floop}, which preserves both $U(1)$ and $SU(2)$ 
Ward identities. Recently, this scheme has been generalized to the case 
of massive external fermions, both in its minimal version, which 
considers the imaginary parts of the fermionic loops (\ifl)~\cite{ifl}, 
and in its full realization with real and imaginary 
parts~\cite{rfl0,rfl1}. In 
particular, in ref.~\cite{ifl} a detailed numerical investigation has been performed, 
showing no significant difference between the \ifl and the 
fixed width scheme, even in the region most sensible to $U(1)$ 
gauge invariance. For this reason, the fixed width scheme is adopted 
in the present calculation.

Another delicate issue is the so called resolved-photon component of the cross
section. The quasi-real $t$-channel photon can split into a pair of almost 
massless quarks, leading to a situation where the 
partonic picture of hadrons is inadequate and both perturbative and 
non-perturbative \qcd corrections become relevant. 
This issue is widely discussed in the literature~\cite{stein}, 
where the standard approach to this problem is also 
described and to which the reader is referred for details. 
However, for single-$W$ like events the resolved-photon component
does not constitute a severe limitation: once a hard $q{\bar q}$
invariant mass cut is imposed, as done in realistic 
data analysis, the bulk of the signal is kept, whereas
the resolved photon contribution becomes almost 
negligible~\cite{passarino}. 
 
A further relevant issue is given by 
radiative corrections due to photon radiation. 
Because exact $O(\alpha)$ electroweak corrections to single-$W$ production 
are still unknown, in most of the theoretical and experimental
studies presented insofar 
only the large contribution of initial-state radiation (\isr) has been 
taken into account, generally by using
collinear structure function (hereafter denoted as \csf) and assuming 
$s = 4E^2$ as the proper scale for \qed radiation. Due to the dominance
of the quasi-real $t$-channel photon exchange, this can be expected not to be 
 a suitable choice in the present case. 
On the other hand, it has been recently proposed to
correct only the $s$-channel 
contributions to the single-$W$ signature, fixing the 
radiation scale in the usual manner, and to neglect the 
photonic corrections to the $t$-channel contributions~\cite{dub}. 
Following previous investigations \cite{bhabha,sisr,babayaga,bdk,bbv} of the 
pattern of 
photonic radiation in \qed and electroweak processes, some 
theoretical arguments to determine the appropriate energy scale 
entering the \csf are presented and compared with existing results. 
The analysis here described elucidates the theoretical details and 
provides further numerical results of a contribution by the 
authors~\cite{apt} to the activity
 of the four-fermion working group of the 
\lepb \mc workshop at \cern~\cite{proposal}.
Ideas similar to those adopted in the present work 
have recently appeared in ref.~\cite{kuriharat} 
and there applied to  the two-photon 
process $e^+ e^- \to  e^+ e^- \mu^+ \mu^-$.

The paper is organized as follows. After a short review 
of the \csf approach to leading log (\lls) \qed corrections in Sect.~\ref{brems}, 
Sect.~\ref{analytic} collects the analytical results 
valid for soft and collinear corrections to a 
generic scattering process. By comparing the 
results of  Sect.~\ref{brems} and 
Sect.~\ref{analytic}, the radiation scales for single-$W$ 
 production are determined in Sect.~\ref{scale}. 
Sect.~\ref{alphar} deals 
with the problem of taking into account the effect of the photon
vacuum polarization in the single-$W$ process, while 
Sect.~\ref{numeric} 
shows the numerical results of 
the present study obtained with a Monte Carlo 
(hereafter \mc) code for the single-$W$ 
signature, including 
also the 
effect of anomalous trilinear gauge couplings. 
Conclusions and prospects
are given in Sect.~\ref{ending}.

\section{\label{brems} Structure Function approach 
to photon radiation}

Since in high-energy processes the corrections due to the emission 
of soft and collinear radiation are quite large, 
the \lls contribution
must be calculated at every perturbative order. A common technique 
to achieve this goal is the \qed structure
function approach \cite{sf}, which consists in convoluting the 
hard-scattering cross section with appropriate 
``parton'' densities. As well known, these convolution factors, 
{\it i.e.} the \qed structure functions,
include, by construction, both the real and virtual part of the photon 
correction, in order to ensure the cancellation of the 
infrared singularities. If a generic Born-level
prediction $d\sigma_0$ is considered, the cross section 
$d\sigma$ including \lls \qed radiative corrections is obtained, by 
virtue of factorization theorems, 
according to the following general formula~\cite{sf}
\begin{equation}
\label{eq:sf}
 d \sigma = \prod_i \int  dx_i D(\Lambda^2,x_i) \; d \sigma_0
\end{equation}

\noindent where $1-x_i$ are the energy fractions carried away by the radiated
photons from the $i$-th leg, $\Lambda$ is the characteristic scale 
of the \csf
$D(\Lambda^2,x_i)$, whose evolution is driven by the 
Dokshitzer-Gribov-Lipatov-Altarelli-Parisi 
(DGLAP) equation~\cite{dglap} and is dependent on $\Lambda$. It is worth 
noticing that the choice of the scale $\Lambda$ is not
dictated by general arguments and it is therefore
rather arbitrary. 

Equation (\ref{eq:sf}) can be rewritten by stressing the possibility 
of different scales for each \csf as follows

\begin{equation}
\label{eq:sfgen}
 d \sigma = \prod_i \int dx_i D(\Lambda_i^2,x_i) \; d \sigma_0
\end{equation}

In particular, if the integrated 
hard-scattering cross section is a smooth function 
of the centre of mass (c.m.) energy, once the integrations
over the energy fractions $x_i$ are performed 
in the soft-photon approximation,
the $O(\alpha)$ double-log expansion of eq. (\ref{eq:sfgen}) 
can be written as follows

\begin{equation}
\label{eq:lle}
  d \sigma = d \sigma_0 \left ( 1 + \sum_i \frac{\alpha}{\pi} 
\log \frac{\Delta E}{E} L(\Lambda_i^2) \right )
\end{equation}
where $\Delta E/ E$ is 
the maximum energy for undetected photons, to be identified with 
finite energy resolution of the photon detector, and 
$L (\Lambda_i^2) \equiv \log(\Lambda_i^2/m^2)$ is the 
collinear logarithm.

Since the functional form of the \qed \csf is accurately known~\cite{sf}, 
the main problem in evaluating eq.~(\ref{eq:sfgen}) is to
fix the process scales $\Lambda_i$. A 
generally adopted attitude is given by the comparison of eq.
(\ref{eq:lle}) with a perturbative calculation, which can be performed within any
approximation, provided it reproduces the correct  
double-log contribution of the $O(\alpha)$ correction. 
This issue is addressed in the next Section.

\section{\label{analytic} Analytical results}

The double-log contribution to photon radiation 
traces back to soft and collinear 
bremsstrahlung and its virtual counterpart~\cite{yfs}, and, in the 
case of a calorimetric measurement 
of the energy of the final-state (\fs) particles, to hard radiation 
collinear to the \fs particles themselves~\cite{calo,gatto,jj,cacci}. 
At each 
perturbative order, the leading contribution can be expanded  in terms
of infrared and collinear logs. For example, when the 
 photons are emitted from the initial-state (\is)
particles only, such an expansion can be arranged in terms of 
double-log contributions of the form 
$\alpha^n l^n L_{s}^n$, where 
$l \equiv \log (\Delta E/E)$ is the infrared log  
and $L_{s} \equiv \log (s/m^2)$ is the collinear log. This is 
the reason why $\Lambda^2 = s$ is the ``natural'' energy scale to be 
used for \csf in the presence of \isr only.
When also \fs radiation is considered, the collinear log, $L$,
is in general modified by additional factors coming from the angular integration 
over the photon variables. A typical example is given by the 
radiation emitted from one leg in the $t$-channel 
\qed contribution to Bhabha scattering~\cite{bhabha}.  
In the soft-photon approximation the radiation cones,
one from the \is electron and one from the \fs electron,
have a half-opening which is determined by the angle 
between the emitting particles,
because of a destructive interference. As a consequence, the 
energy scale $s$, which
appears in $L_{s}$, transforms into $|t| \simeq s(1-\cos\theta)$, where $\theta$ is
the electron scattering angle, and, therefore, $L_{s} \to L_{t}$. 
Hence the perturbative expansion contains collinear logs which are modified
because of the angular ordering introduced by the radiation cones.
In the presence of 
large scattering angles, for which $|t| \simeq s$, the above modification 
is numerically small, but it becomes more and more important 
in the forward angular range, which is the dynamically
favourite region by $t$-channel Bhabha scattering and where $t \ll s$. 
The net result is a numerically significant 
depletion of \qed radiation effects just in the most important 
 part of the hard-scattering $t$-channel dynamics. 
 Actually, when using \csf to evaluate \qed 
 \lls corrections to small-angle Bhabha scattering, the 
 energy scale $\Lambda^2 = |t|$ is employed in 
 all phenomenological applications~\cite{bhabha}. 
More in general, in order to take into account 
dominant initial-final-state interference effects in addition to 
initial- and final-state leading terms, $s$ and $t$ \qed 
contributions to Bhabha scattering 
can be corrected in terms of a unique combination 
of Mandelstam invariants given by $s t / u$, 
 as discussed in refs.~\cite{sisr,babayaga}. Therefore, the energy scale 
 $\Lambda^2 = s t/u$ turns out to be a suitable choice for the 
 evaluation in terms of \csf of LL corrections to \qed Bhabha scattering, 
 as demonstrated, in comparison with the exact $O(\alpha)$ calculation, 
 in ref.~\cite{babayaga}. Similar arguments for an appropriate 
 choice of the energy scale for QED radiation, based on the inspection of the 
 soft and collinear limit of the $O(\alpha)$ correction, have been 
 also advocated in ref.~\cite{bdk} for the reaction 
 $e^+ e^- \to W^+ W^-$ and, very recently, in ref.~\cite{kuriharat} for 
 the process $e^+ e^- \to  e^+ e^- \mu^+ \mu^-$. Comparisons 
 performed in refs.~\cite{bdk,kuriharat} with available 
 exact $O(\alpha)$ calculations explicitly exhibit the 
 validity of such a strategy, which is therefore pursued 
 in the present analysis. As already remarked, 
 the result for \lls corrections in the presence 
 of a calorimetric detection of \fs particles must include the
contribution of photons which, regardless of their energy,  
can not be discriminated from closely collinear 
fermions, as a consequence of the finite angular resolution 
of the calorimeters.
The role of such hard photons collinear 
to the \fs particles becomes therefore 
unavoidable in the case of a calorimetric 
measurement of the energy of the \fs particles, 
as discussed in the following.

\subsection{\label{soft} Soft-photon contribution}

In this Section the contribution of photons, too soft to be 
detected in the calorimeter, 
will be computed for a generic process 
with $n$ ingoing-legs. \footnote{\label{legs}
This choice fixes our conventions. Outgoing particles will appear as 
ingoing ones with momentum and charge according to crossing symmetry.} 
The following approximations are understood 

\begin{equation}
  \label{eq:spa}
  \left \{ \begin{array}{ll}
    q_i \gg k \\
    s_{ij} \gg m_i^2,\,m_j^2 \\
  \end{array} \right .
\end{equation}

\noindent where $q_i$ is the momentum of a particle of mass $m_i$ emitting
a real photon of momentum $k$, and $s_{ij}\equiv(q_i+q_j)^2$ is the 
invariant mass of the pair $ij$.

By following the standard derivation of the eikonal 
factors due to soft bremsstrahlung and
by generalizing it to particles with different masses and charges, the 
differential cross section, dressed by soft-photon emission,  
can be cast into the following factorized 
form~\cite{yfs}

\begin{equation}
\label{eq:soft}
  d\sigma_{\rm soft} = d\sigma_0 \frac{d\omega}{\omega} \frac{2\alpha}{\pi}
\sum_{i>j}^n Q_iQ_j \log \frac{s_{ij}}{m_im_j}
\end{equation}

\noindent where $\omega$ is the photon energy and $Q_i$ is the charge of
the $i$-th particle.

It is worth noticing that in the limit $s_{ij} \ll m_i^2,\,m_j^2$, provided
the first inequality in eq.~(\ref{eq:spa}) still holds, the logarithmic behaviour
present in eq.~(\ref{eq:soft}) disappears, leaving a power law 
which can be simply obtained by means of the 
substitution~\cite{yfs}

\begin{equation}
\label{eq:pancheri}
  \log \frac{s_{ij}}{m_im_j} \longrightarrow \frac{1}{3} \frac{s_{ij}}{m_im_j}
\end{equation}

Notice that, since the goal is to 
determine the scale entering the \csf, only the contribution 
of real photons is explicitly calculated, because the 
virtual corrections, in order to preserve the
cancellations of infrared 
singularities, must share the same leading collinear structure of the real
part itself.

By including the 
virtual part needed to cancel the 
infrared singularity and 
integrating eq. (\ref{eq:soft}) over the photon energy $\omega$ 
in the soft region $0 \leq \omega \leq \Delta E$, one 
gets 
\begin{equation}
\label{eq:sv}
  d\sigma_{\rm S+V} = d\sigma_0 
\log \frac{\Delta E}{E} \frac{2\alpha}{\pi}
\sum_{i>j}^n Q_iQ_j \log \frac{s_{ij}}{m_im_j}
\end{equation}

\subsection{\label{hard} Hard radiation collinear to the final-state particles}

In the case of a calorimetric set-up, which is the 
realistic situation for single-$W$ production at \lep, photons 
collinear to the detected 
\fs particles can not be distinguished from the 
the emitting particles themselves, independently of the 
photon energy. Therefore, in order to 
obtain the correct structure of double-log corrections 
for such an event selection, the effect due to the emission of unresolved 
hard radiation collinear to the \fs particles must be 
taken into account in addition to soft+virtual corrections.    

To this end, let us 
re-consider the previous process with $n$ ingoing legs and think $m$ of them
to be changed into outgoing legs at the end of the calculation (see 
the previous footnote). Then, 
the contribution of photons collinear to the 
\fs particles can be cast into
a gauge invariant form as follows~\cite{gatto,jj,cacci}

\begin{equation}
\label{eq:hard}
  d\sigma_{\rm hard} = d\sigma_0 \frac{d\omega}{\omega} \frac{2\alpha}{\pi}
\sum_i^m Q_i^2 \log \frac{E_i\delta}{m_i}
\end{equation}
              
\noindent where $E_i$ is the energy of the $i$-th particle, and $\delta$ is the
half-opening angle of the calorimetric resolution. 

By integrating eq. (\ref{eq:hard}) over the photon energy $\omega$ 
in the range $\Delta E \leq \omega \leq E$, the integrated 
correction due to hard photons collinear to the
\fs particles is given by  
\begin{equation}
\label{eq:hardi}
  d\sigma_{\rm hard} = d\sigma_0 
\log \frac{E}{\Delta E} \frac{2\alpha}{\pi}
\sum_i^m Q_i^2 \log \frac{E_i\delta}{m_i}
\end{equation}

\subsection{\label{master} The master formula}

Equations (\ref{eq:sv}) and (\ref{eq:hardi}) give the leading 
double-log contribution which
must be compared to (\ref{eq:lle}), the $O(\alpha)$ perturbative expansion of
eq.~(\ref{eq:sfgen}), in order to fix the process scales $\Lambda_i$.
Summing the contributions of eq. (\ref{eq:sv}) and eq.~(\ref{eq:hardi}), the 
analytical cross section is in conclusion given by

\begin{eqnarray}
\label{eq:master}
  d\sigma_{\rm S+V} + d\sigma_{\rm hard} & = &
  d\sigma_0 \frac{2\alpha}{\pi} \log \frac{\Delta E}{E}
  \left \{ \sum_{i=m+1}^n Q_i^2
\log \frac{E_i}{m_i} 
 \; + \right . \\ \nonumber
  & & - \; \left . \sum_{i>j}^n Q_iQ_j \log 2(1-c_{ij}) -
    \sum_i^m Q_i^2 \log \delta \right \}
\end{eqnarray}

\noindent where $c_{ij}$ is the cosine of the angle between particles $i$ and $j$.

Three different 
kinds of logarithms occur in eq.~(\ref{eq:master}). The first term
contains the mass and energy logarithms of the \is particles 
only, since, as expected, the energies and the masses of the \fs 
particles disappeared, in agreement with
the {\sc kln} theorem \cite{kln}. The second term includes angular terms 
due to radiation interference, while the third one comes from the 
requirement of calorimetric 
measurement.

These terms must be compared with the collinear logarithms of eq.~(\ref{eq:lle})
in order to fix the scales $\Lambda_i$ of the \csf. In the following 
Section
this task is accomplished in detail for the single-$W$ process.

\section{\label{scale} Fixing the radiation scales in the single-$W$ process}

Let us consider, for definiteness, 
the process $e^+e^- \rightarrow e^-{\bar\nu}u{\bar d}$
with the \fs electron lost in the beam pipe 
(single-$W$ process). In this event selection
(hereafter \es) the leading contribution comes from $\gamma^* e^+$ scattering
with the virtual photon emitted from the electron line. 
The leading dynamics is given by the
$t$-channel Feynman diagrams shown in Fig.~\ref{fusion}.

\begin{figure}[h]
\begin{center}
  \includegraphics[bb=135 605 290 720,scale=1.]{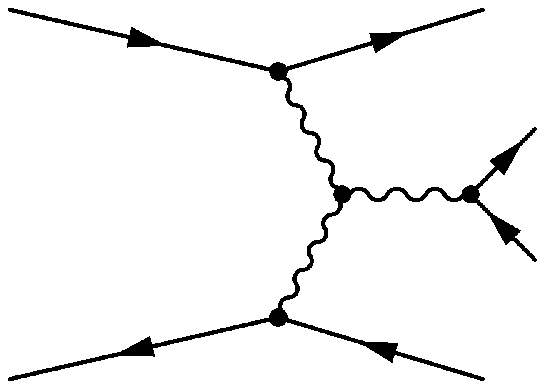}
  \hspace*{2em}
  \includegraphics[bb=135 605 290 720,scale=1.]{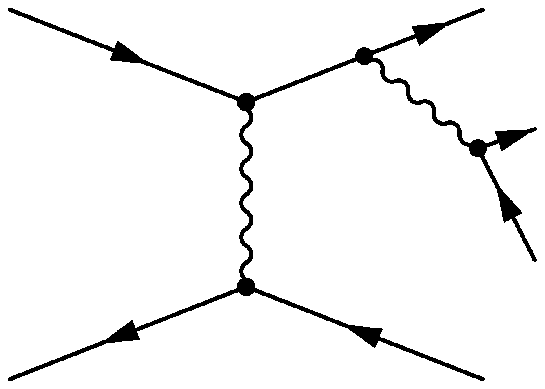}
\end{center}
\caption[fusion]{\label{fusion}
  The fusion and bremsstrahlung diagrams are the leading Feynman graphs
  for the single-$W$ signature.}
\end{figure}

If a calorimetric measurement of the energies of 
the \fs particles is 
performed, only the \is legs need to be corrected by
the \csf. Furthermore, since the electron is scattered in the very
forward region, the interference between the electron line and the rest of the
process is very small. This allows a natural sharing of the logarithms coming from
eq. (\ref{eq:master}) between the two \is \csf 
associated to the colliding electron and positron.

Hence the formula (\ref{eq:master}),  
when compared with eq. (\ref{eq:lle}), 
translates into the two following scales
($\Lambda_-$  refers to \csf 
attached to the electron line, while $\Lambda_+$ to 
the \csf attached to the positron line),

\begin{equation}
\label{eq:wscales}
  \Lambda_-^2 = 4E^2 \frac{(1-c_-)^2}{\delta^2} \;,\quad
  \Lambda_+^2 = 2^{\frac{14}{9}}E^2
          \frac{\big((1-c_{\bar d})(1-c_u)^2\big)^\frac{2}{3}}
          {\big((1-c_{u{\bar d}})^2\delta^5\big)^\frac{2}{9}}
\end{equation}

\noindent where $E$ is the beam energy, $c_-$ is the cosine
of the electron scattering angle, $c_u$ and $c_{\bar d}$ are 
the cosine of the quark
scattering angles with respect to the initial positron, 
and $c_{u{\bar d}}$ is the cosine
of the relative angle between the quarks.

It is worth noticing that in the numerical implementation, whenever one of the
two scales is less than a small cut-off 
($\Lambda^2_{\rm cut-off} = 4 m_e^2$, where 
$m_e$ is the electron mass), the 
radiation from the corresponding leg is switched off, in accordance with 
eq. (\ref{eq:pancheri}). 
It was carefully 
tested that variations of the cut-off do not alter the numerical 
results.

Owing to the presence of a resonant $W$ boson, some modifications to the previous
results may come from finite-width effects and from 
radiation decoherence~\cite{mg}.
Finite-width corrections of the form of $E_\gamma/\Gamma_W$ arise when the
unstable particle propagator goes off its mass-shell, but this is not the present
case, since the multi-fermion final state can accommodate a resonant $W$.
Radiation decoherence is present whenever a resonance occurs and its effect is to
cancel the angular dependence from the scale. As a consequence the scale $\Lambda_+$
should be modified by dropping the angular interference factors in eq.
(\ref{eq:wscales}) when the emitted photons have $E_\gamma \sim \Gamma_W$.
Yet in the present case the effect is tiny, since the effects due to angular
interference for the scale $\Lambda_+$ are already small by themselves.

It is also possible to make a naive ansatz for the 
radiation scales without a detailed
calculation, by thinking of the graphs of Fig.~\ref{fusion} in terms of the
Weizs\"acker-Williams approximation \cite{ww}, {\it i.e.} in terms of a convolution
of the process $e^+\gamma \rightarrow \nu_e W^*$ with an equivalent
photon spectrum plus a real electron line. This leads to 
assigning two different
scales to the single-$W$ process: one scale for the electron current and one for
the positron current. The former scale is the proper one for a 
$t$-channel process,
e.g. $t$-channel Bhabha scattering, so it is simply 
$|q^2_{\gamma^*}|$, where 
$|q^2_{\gamma^*}|$ is the squared momentum transfer 
in the $ee\gamma^*$ vertex. The latter is the sum of
an $s$-channel electron exchange and a $t$-channel $W$ exchange (see Fig.
\ref{fusion}). Assuming that the $t$-channel dominates, its natural cut-off is given
by the $W$-boson mass, $M_W$. Hence, the following ansatz follows

\begin{equation}
\label{eq:naive}
  \Lambda_{-,{\rm naive}}^2 = |q^2_{\gamma^*}| \;,\quad
  \Lambda_{+,{\rm naive}}^2 = M_W^2
\end{equation}

\noindent where $M_W$ is the mass of the $W$ boson.
The comparison between the scales given by eq. (\ref{eq:wscales}) and
these naive scales, which will be performed numerically in the following 
Section,
provides a useful cross-check of the analytical results 
derived by inspection with the soft/collinear limit of 
the $O(\alpha)$ correction.

A discussion of other possible approaches to the treatment 
of photonic corrections to single-$W$ production can be 
found in the four-fermion working group 
report of the \lepb \mc Workshop~\cite{proposal}.

\section{\label{alphar} The running of the electromagnetic coupling 
constant}

Besides the higher-order \qed corrections discussed in the 
previous Sections, other large logarithmic contributions 
to the single-$W$ cross section 
arise from the running of the electromagnetic coupling 
constant $\alpha$. Since  in the case under study the dominant 
configurations come from the Feynman 
diagrams with an almost on-shell photon exchange,  
the appropriate scale for the evaluation of the electromagnetic 
coupling relative to the $t$-channel 
photon in the $ee\gamma^*$ vertex is the squared momentum transfer 
$q^2_{\gamma^*}$ defined above. 

However, because $G_F$, $M_W$ and $M_Z$ 
are the typically adopted input parameters for 
electroweak processes at \lepb, the electromagnetic 
coupling is fixed at tree-level to a high energy value as, for example,
\begin{equation}
\alpha_{G_F} = 4 \sqrt{2} {{G_F M_W^2 s_W^2} \over {4 \pi}}\;, \quad \quad {\rm with} 
\quad \quad 
s_W^2 = 1 - {{M_W^2}\over {M_Z^2}} \;.
\end{equation}

On the other hand, the single-$W$ production is a 
$q^2_{\gamma^*} \simeq 0$ dominated process and therefore 
the above high-energy evaluation of $\alpha$, $\alpha_{G_F}$, 
needs to be rescaled to its correct value at small momentum 
transfer. To this end,  a gauge-invariant ``reweighting''
procedure can be adopted, by rescaling the differential cross section 
$d\sigma/dt$ ($t \equiv q^2_{\gamma^*}$) in the 
following way

\begin{equation}
\label{eq:running}
{{d\sigma}\over{dt}} \rightarrow {{\alpha^2(t)}\over{\alpha^2_{G_F}}}
{{d\sigma}\over{dt}} \;,
\end{equation}
where $\alpha(t)$ is the running coupling 
constant computed at virtuality  $q^2_{\gamma^*}$.  

A detailed analysis 
of the effect of the running couplings in single-$W$ production 
has been recently performed 
within the massive fermion-loop scheme in ref.~\cite{rfl1}, where the couplings 
are automatically running in the calculation. 
As shown in ref.~\cite{rfl1}, the relative difference  
between the above reweighting prescription and the complete results 
of the fermion-loop scheme is at the 1\%-2\% level~\footnote{
Actually, for the single-$W$ 
final state under examination here and for realistic event selections, 
the differences between the two procedures 
are confined below the 1\% level.}, and it is 
therefore in the expected range of theoretical uncertainty 
due to missing full one-loop electroweak corrections.

\section{\label{numeric} Numerical results}

In this Section the \mc code, developed to simulate the single-$W$
process, is described and a sample of numerical results 
obtained by means of it is shown and commented, 
with particular emphasis on the effects of higher-order \qed 
corrections to single-$W$ production at \lepb energies. 

\subsection{\label{mc} The Monte Carlo code}

A \mc program, named {\tt SWAP}, was developed to calculate 
cross sections and differential 
distributions for the single-$W$ signature. 

As already emphasized, the main feature of this process is the fact 
that the $t$-channel photon of Fig.~\ref{fusion} becomes quasi-real.
In the limit of massless fermions, 
the photon propagator becomes singular in the
forward direction and the cross section develops a logarithmic singularity.
Indeed, whenever the final electron is lost in the beam pipe, its mass becomes
a natural cutoff for the very-forward singularities, compelling to build a
massive matrix element and phase-space. 
The phase-space integration is performed in {\tt SWAP} with the aid of a
multi-channel importance sampling with stratification. The main
peaking structures for the single-$W$ process 
are given by the dynamics depicted by the fusion and bremsstrahlung
graphs of Fig.~\ref{fusion}. They are the resonant $W$-boson
invariant mass, treated with a Breit-Wigner weight,
and the $t$-channel ``singularity'' of the quasi-real photon, treated
with a $1/|t|$ weight.
Moreover, the program can deal with the singularities of the sub-leading
$t$-channel CC$20$ diagrams shown in Fig.~\ref{other}, by means
of the multi-channel approach.

\begin{figure}[h]
\begin{center}
  \includegraphics[bb=135 605 290 720,scale=1.]{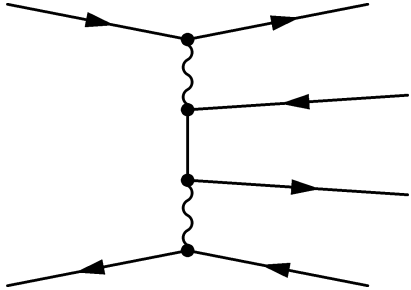}
\end{center}
\caption[fusion]{\label{other}
  The multiperipheral diagram is the main sub-leading Feynman
  graph for the single-$W$ signature.}
\end{figure}

The exact hard-scattering 
matrix element is computed by means of the \alp code \cite{alpha} for
the automatic evaluation of Born scattering amplitudes. Fermion
masses are exactly taken into account and the fixed-width scheme is 
adopted as gauge-restoring approach, 
by taking the massive gauge boson propagator as follows:
\begin{equation}
\Pi^{\mu \nu} = {-i\left( g^{\mu\nu}-{k^\mu k^\nu\over M^2 - 
    i\Gamma M}\right)\over
    k^2 - M^2 + i\Gamma M} \,,\quad\Gamma = {\rm ~cost.}
\end{equation}
It is known~\cite{ifl,fwidth,floop} that this scheme preserves 
$U(1)$ gauge invariance but still violates
$SU(2)$ Ward identities. However, at least in the unitary gauge employed here,
it is indistinguishable from other fully gauge-invariant 
schemes \cite{ifl,fwidth,floop}.

The contribution of anomalous gauge couplings
is also accounted for in {\tt SWAP}. The anomalous gauge boson couplings $\Delta k_{\gamma}$, $\lambda_{\gamma}$, 
$\delta_Z$, $\Delta k_Z$ and $\lambda_{\gamma}$ are implemented in the 
{\tt ALPHA} code according to the
parameterization of refs.~\cite{gg,hhpz}. Photon radiation
is implemented via \csf formalism, according
to the discussion of Sect.~\ref{scale}. 
It is worth noticing that, since the incoming electron/positron 
are required to be on-shell massive 
fermions, a naive four-momentum rescaling, due to
photon emission, such as $\hat{p}_{\pm} = x \, p_{\pm}$ leads to 
potentially dangerous gauge violations, according to what 
previously discussed. Therefore, the rescaled incoming four-momenta 
are implemented as $\hat{p}_{\pm} = (x \, E, 
0, 0 , \pm \sqrt{x^2 E^2 - m_e^2})$, by interpreting $x$ as the 
energy fraction after photon radiation, as motivated in 
ref.~\cite{babayaga}. If required, $p_\perp / p_L$ effects can be 
provided in the treatment of ISR, by means of either $p_\perp$-dependent 
SF~\cite{nunugpv} or a QED 
Parton Shower algorithm~\cite{babayaga,jap}.  
The effect of vacuum polarization is 
taken into account as described by 
eq.~(\ref{eq:running}), by including the 
contribution of leptons, heavy quarks and light 
quarks, the latter according to the parameterization of 
ref. \cite{vpol}. The program supports 
realistic \es and it can be employed either as
a cross-section calculator or as a event generator, with both weighted and
unweighted generation available.

The technical precision of the event generator {\tt SWAP} has already been 
carefully proved in ref.~\cite{proposal}, by means 
of detailed tuned comparisons between the predictions 
of independent codes. Perfect agreement was found, both 
at the level of integrated cross sections and distributions, 
also for purely leptonic final states.

\subsection{\label{numdisc} Discussion of the numerical results}

\begin{table}
\caption[calo]{\label{calo}
  The \es adopted for the calculations shown in the present paper
  for the signature $e^+ e^- \rightarrow e^- \bar \nu_e u \bar d$,
  according to ref.~\cite{proposal}.} 
\medskip
\begin{center}
\begin{tabular}{|l||c|c|} \hline
 electron angular acceptance      &  $|\cos\theta_e| > 0.997$ &  
 $|\cos\theta_e| > 0.997$  \\
 quarks angular acceptance        &  1. no cut  &  
 2. $|\cos\theta_{q, {\bar q}}| < 0.95$  \\
 calorimetric half-opening angle  
 &  $5.00^\circ$ & $5.00^\circ$     \\ \hline
 quark-antiquark invariant mass &  $45.0~{\rm GeV}$ & 
 $45.0~{\rm GeV}$ \\ \hline
\end{tabular}
\end{center}
\end{table}

The numerical simulations are elaborated
according to the \es reviewed in Tab.~\ref{calo}, with
the electroweak input parameters shown in Tab.~\ref{scheme}.

\begin{table}
\caption[calo]{\label{scheme}
  The adopted electroweak input parameters, 
  according to ref.~\cite{proposal}. 
  All other parameters are
  calculated by means of the tree-level relations.}
\medskip
\begin{center}
\begin{tabular}{|l|} \hline
                $G_F = 1.16637 \times 10^{-5}~{\rm GeV}^{-2}$ \\
                $M_Z = 91.1867~{\rm GeV}$ \\
                $M_W = 80.35~{\rm GeV}$ \\ \hline
                $\Gamma_Z = 2.49471~{\rm GeV}$ \\ \hline
\end{tabular}
\end{center}
\end{table}

\begin{figure}
\begin{center}
\includegraphics[bb = 40 380 525 610, scale = .8]{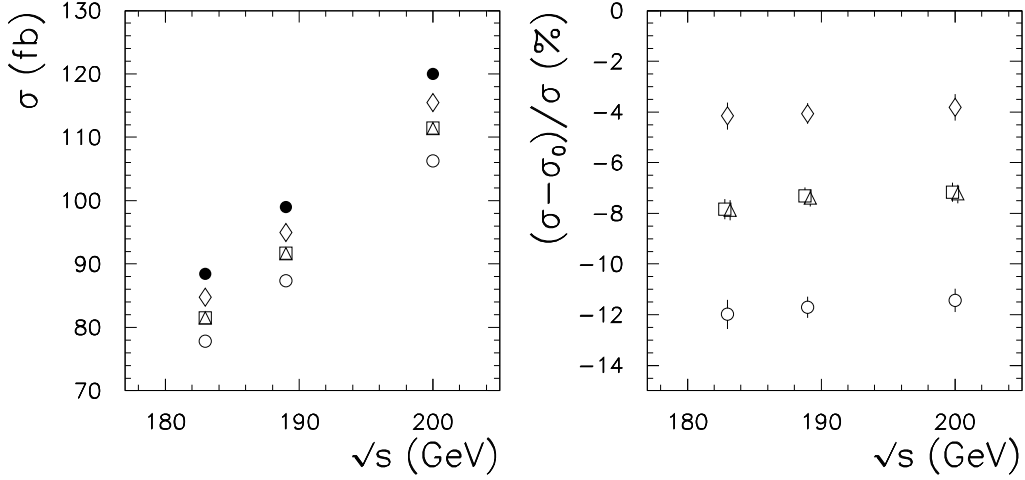}
\end{center}
\caption{The effects of \lls \qed corrections to the cross section 
         of the single-$W$ process
         $e^+e^-\rightarrow e^-\bar{\nu}u\bar{d}$ for different choices 
         of the energy scale in the electron/positron \csf. 
         The quark angular 
         acceptance $0^\circ \leq \vartheta_{u,\bar d} \leq 180^\circ$
         is considered.
         Left: 
         absolute cross values as functions of the \lepb 
         c.m. energy. Right: relative difference between the \qed 
         corrected cross sections and the Born one, still as 
         functions of the c.m. energy. The marker
         $\bullet$ represents the Born cross section,
         $\scriptstyle\bigcirc$ represents the correction
         given by $\Lambda^2_{\pm} = s$ for both \csf,
         $\scriptstyle\diamondsuit$ the correction given
         by the scales $\Lambda^2_{\pm} = |q^2_{\gamma^*}|$ 
         for both \csf, 
         $\scriptstyle\square$ the correction
         given by the naive scales of eq.~(\ref{eq:naive}),
         $\scriptstyle\triangle$ the correction given by the 
	 scales of eq.~(\ref{eq:wscales}).
         The entries correspond to 
         $\sqrt{s} = 183$, $189$, $200$ ~GeV.
         The markers are misplaced for better reading.}
\label{varying}
\end{figure}

\begin{figure}
\begin{center}
\includegraphics[bb = 40 380 525 610, scale = .8]{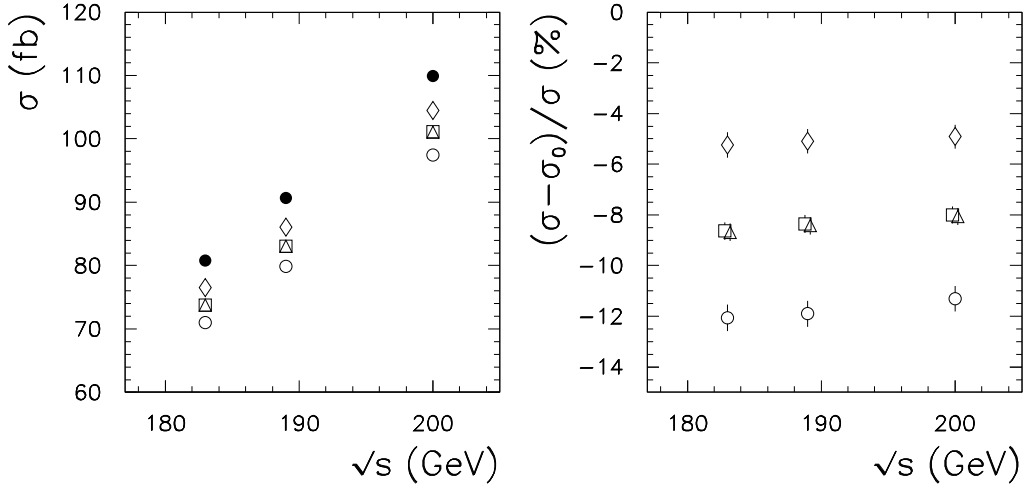}
\end{center}
\caption{The same as Fig.~\ref{varying} for the quark angular acceptance 
$|\cos\theta_{u, {\bar d}}| < 0.95$.}
\label{varying1}
\end{figure}

In Figs.~\ref{varying}-\ref{varying1} the numerical impact of different 
choices of the $\Lambda^2$-scale on the 
cross section of the single-$W$ process
$e^+e^-\rightarrow e^-\bar{\nu}u\bar{d}$ 
in the \lepb energy range is shown. Since the 
energy scale $\Lambda_+$ of eq.~(\ref{eq:wscales}) depends
on the quark scattering angles, two different quark angular 
acceptances are considered, namely no cut  
(Fig.~\ref{varying}) and  $|\cos\vartheta_{u, {\bar d}}| < 0.95$ 
 (Fig.~\ref{varying1}).  
The marker  $\bullet$ represents the Born cross section,
$\scriptstyle\bigcirc$ represents the correction
given by $\Lambda^2_{\pm} = s$ scale for both IS \csf, 
$\scriptstyle\diamondsuit$ represents the correction
given by $\Lambda^2_{\pm} = |q^2_{\gamma^*}|$ scale for both IS \csf,
$\scriptstyle\triangle$ the correction given by the scales 
of eq.~(\ref{eq:wscales}),
$\scriptstyle\square$ the correction
given by the naive scales of eq.~(\ref{eq:naive}). 
It can be seen that neither the 
$s$ scale, as implemented in some computational tools used 
for the analysis of the single-$W$ process, nor the 
$|q^2_{\gamma^*}|$ 
scale are able to reproduce the effects due to the scales 
of eq.~(\ref{eq:wscales}) and eq.~(\ref{eq:naive}). These 
two scales are in good agreement and both predict a 
lowering of the Born cross section of about 8-9\%, almost 
independent of the c.m. \lepb energy 
and quark angular acceptance.
This fact can be understood by looking at Fig.~\ref{scales}, 
where it is shown the
single-$W$ differential cross section with respect to the scales 
$\Lambda_{\pm}$ of eq.
(\ref{eq:wscales}). On the left, $\Lambda_{+}$ exhibits 
a broad peak not far from $M_W$, while, on the 
right, the other
scale $\Lambda_{-}$ peaks, as expected, at very small 
momentum transfer.

\begin{figure}
\begin{center}
\includegraphics[bb=40 380 525 610,scale=.8]{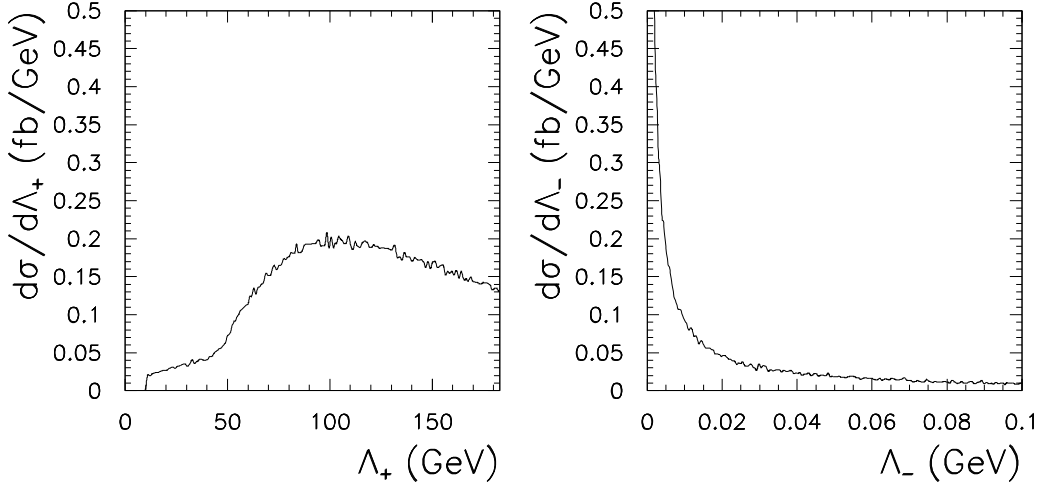}
\end{center}
\caption[fusion]{\label{scales}
  The differential cross sections of the 
single-$W$ process $e^+e^-\rightarrow e^-\bar{\nu}u\bar{d}$ 
with respect to the two scales of eq. (\ref{eq:wscales}) 
at $\sqrt{s} = 189$ GeV.}
\end{figure}

Figure~\ref{figal} 
shows the effects of the reweighting procedure of eq. (\ref{eq:running}) 
for the evaluation of the \qed running coupling constant. 
The marker $\scriptstyle\triangle$ represents the relative 
difference between the integrated cross section computed 
in terms of $\alpha_{G_F}$ and the cross section 
computed in terms of $\alpha(0)$, while the marker
$\scriptstyle\diamondsuit$ is the relative 
difference between the integrated cross section computed 
in terms of $\alpha_{G_F}$ and the cross section 
computed in terms of $\alpha(t)$. 
As can be seen, the rescaling from $\alpha_{G_F}$ to $\alpha(t)$ 
introduces a negative correction of about 5-6\% in the \lepb 
energy range.
The difference between 
$\scriptstyle\triangle$ and $\scriptstyle\diamondsuit$, 
which is about 2-3\%,  
is a measure of the running of $\alpha_{QED}$ from 
$q^2_{\gamma^*} = 0$ to $q^2_{\gamma^*} = t$.

\begin{figure}
\begin{center}
\includegraphics[bb=-70 400 500 600,scale=.8]{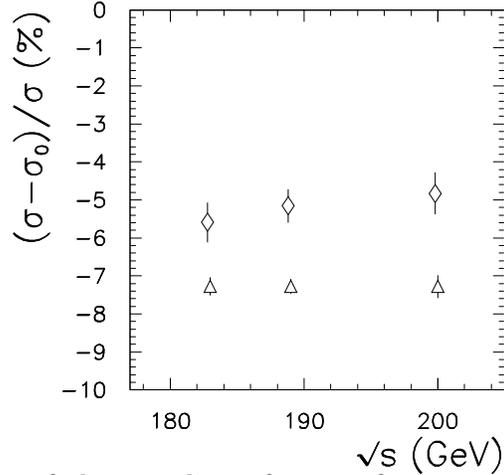}
\caption{The effects of the rescaling of 
$\alpha_{QED}$ from $\alpha_{G_F}$ to $\alpha(q^2_{\gamma^*} = 0)$ 
($\scriptstyle\triangle$) and $\alpha(q^2_{\gamma^*})$  
($\scriptstyle\diamondsuit$)
on the integrated cross 
section of the single-$W$ process
$e^+e^-\rightarrow e^-\bar{\nu}u\bar{d}$. $\sigma_0$ 
is the cross section computed in terms of $\alpha_{G_F}$.
The entries correspond to 
$\sqrt{s} = 183, 189, 200$~GeV. }
\label{figal}
\end{center}
\end{figure}

As an illustrative example of the effect of anomalous couplings 
on single-$W$ differential distributions, in Fig.\ref{mqq} the 
distribution of the $q{\bar q}$ invariant mass, 
around the peak of the $W$-boson resonance, 
and the distribution of the angle of the quarks with the
line of flight of the reconstructed $W$-boson in the $W$-boson
rest-frame are shown. The dashed lines correspond to 
the simulation as obtained by means of {\tt SWAP} for  
 the anomalous coupling 
$\Delta\kappa_\gamma = 0.1$, while the solid lines represent the Standard Model
prediction. The effect of the anomalous coupling 
$\Delta\kappa_\gamma$ 
at \lepb energies is just
an overall rescaling of the total cross section. Therefore the \lepb sensitivity
to $\Delta\kappa_\gamma$ in single-$W$ events depends crucially on the accuracy of the
theoretical evaluation of the total cross section.

\begin{figure}
\begin{center}
\includegraphics[bb=40 380 525 610,scale=.8]{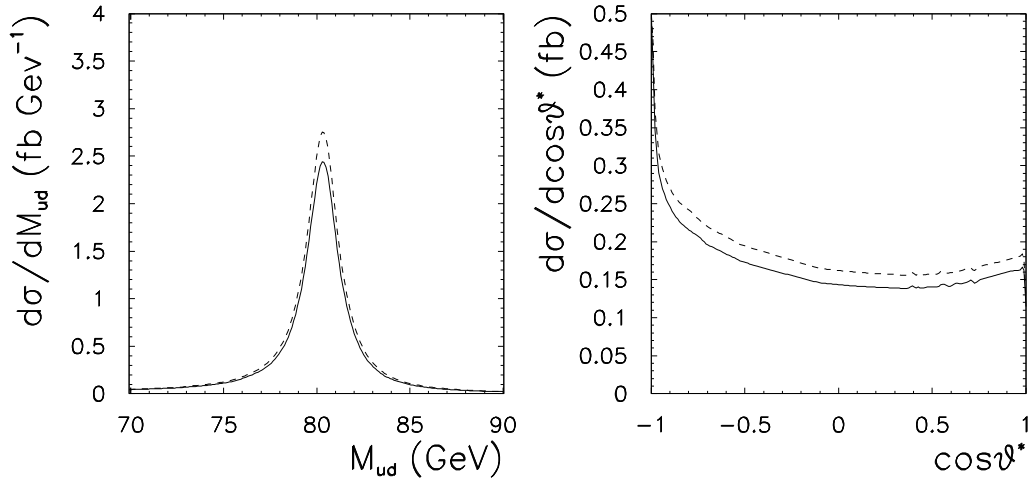}
\end{center}
\caption[fusion]{\label{mqq}
  The single-$W$ differential cross sections with respect to the quark-antiquark
  invariant mass (left side), and with respect to the angle between
   a quark and the line of flight of the reconstructed 
   $W$-boson in the frame of rest of the $W$-boson  (right side).
  The dashed line represents the distribution in the presence of an anomalous
  gauge coupling $\Delta\kappa_\gamma=0.1$, while 
  the solid line is the Standard
  Model prediction. The c.m. energy is $\sqrt{s}$ = 189 GeV.}
\end{figure}

\section{\label{ending} Conclusions}

The process of  
single-$W$ production in high-energy $e^+ e^-$ collisions is relevant 
at \lepb for the determination of the non-abelian 
self-couplings of the $W$ boson, and of primary importance at 
future Linear Colliders at the TeV scale, 
its cross section being dominant at very high energies with respect to other 
four-fermion processes.

In order to give a contribution to the reduction of the 
theoretical uncertainty presently associated to the calculation 
of the single-$W$ cross section, the issue of higher-order 
photonic corrections has been carefully investigated  
 within the standard \csf technique. Theoretical and 
phenomenological arguments for the choice of the energy scale
entering the \csf have been proposed. Two possible solutions 
for the scale of \qed radiation have been obtained.   
The former has been derived   
by means of general arguments concerning the soft and collinear 
limit of the $O(\alpha)$ corrections coming from the radiation of external 
legs. The latter, which can be considered as a naive ansatz, has been
driven by thinking of the single-$W$ process in terms of the 
Weizs\"acker-Williams approximation. 

Numerical calculations show that the typically 
adopted choice of the center-of-mass energy
of the reaction, as radiation scale for the process, can lead to over-estimate 
the radiative
correction by a factor of $1.5$, implying an under-estimate of the cross section
of about $4\%$. Also the choice of fixing the scale to 
the momentum transfer $t$ in the $ee\gamma^*$ vertex for both the
\is \csf leads to an under-estimate of the photon correction
of about $4\%$. The difference between the predictions 
given by the two set of scales 
of eq.~(\ref{eq:wscales}) and eq.~(\ref{eq:naive})
 is at the per mille level in the 
\lepb energy range.
Therefore, the naive scales of eq.~(\ref{eq:naive}) 
provide a good ansatz for the energy scale of QED radiation 
in the single-$W$ process, which could be simply
implemented in MC tools for data analysis
and further corroborated by the comparison
with the results of other approaches.
The method here 
described for the energy scale determination in the \csf can be 
simply generalized to 
other four-fermion process dominated by non-annihilation 
channels, such as single-$Z$ production.

In order to provide adequate phenomenological predictions for 
precision experiments, 
also the running of the electromagnetic coupling constant has been 
accounted for in an effective way, {\it i.e.} by rescaling the differential 
cross section for the ratio of the 
electromagnetic coupling constant, 
evaluated at the typical 
scale of the process, to the same coupling evaluated from the input parameters 
according to tree-level relations.
The effect of such rescaling amounts to a negative 
correction of about $5$-$6\%$, in agreement with recent findings~\cite{rfl1}, 
as far as the effect of 
$\alpha_{\rm QED}$ is concerned.

In the light of the experimental precision for the 
single-$W$ process, the corrections considered 
in the present paper are phenomenologically relevant.

According to the 
theoretical approach described in the present paper, 
an original \mc programme {\tt SWAP} has been developed, 
including exact tree-level matrix elements 
with finite fermion masses effects, anomalous couplings, vacuum 
polarization and higher-order \qed corrections. 
The code is available for experimental analysis.

\begin{ack}
  The authors wish to thank the members of the 
  four-fermion working group of the \lepb Monte Carlo 
  Workshop (\cern), in particular Y.~Kurihara, G.~Passarino, R.~Pittau and 
  M.~Verzocchi, for useful discussions on the subject.
  A. Pallavicini is grateful to the {\sc infn}, Sezione di Pavia, for
  having provided computer facilities.
\end{ack}

\end{document}